\def\preprint{1}  

\ifdefined\preprint
  \documentclass[preprint,times,12pt]{elsarticle}
\fi
\ifdefined\wordcount
  \documentclass[final,5p,times,twocolumn]{elsarticle}
\fi
\ifdefined\final
  \documentclass[final,5p,times,twocolumn]{elsarticle}
\fi

\usepackage{graphicx,stfloats}
\usepackage{color}
\usepackage{jabbrv}
\usepackage{titlesec}
\usepackage{textcomp}
\usepackage{threeparttable} 
\usepackage{caption} 
\usepackage{xcolor}
\usepackage{array}
\usepackage{booktabs}
\usepackage{multirow}
\usepackage{hyperref}
\usepackage{dirtytalk}
\usepackage[section]{placeins}
\usepackage[version=4]{mhchem}
\usepackage[margin=1in]{geometry} 
\usepackage[section]{placeins}
\usepackage{setspace}

\usepackage{amsmath}
\usepackage{amssymb}
\usepackage{graphicx}
\usepackage{latexsym}
\usepackage{times}
\usepackage{textcomp}
\usepackage{chemfig}

\biboptions{sort&compress}
\journal{Combustion and Flame}
\begin{document}
\begin{frontmatter}

\title{Vortex Stretching of Non-premixed, Diluted Hydrogen/Oxygen Flamelets}

\author{Wes Hellwig}
\ead{whellwig@uci.edu}
\author{Xian Shi}
\author{William A. Sirignano}

\address{Department of Mechanical and Aerospace Engineering, University of California, Irvine, Irvine CA, 92697}
\cortext[cor1]{Corresponding author: Wes Hellwig}

\begin{abstract}
A three-dimensional flamelet model considering vortex stretching with unitary Lewis number is used to simulate diluted hydrogen-oxygen diffusion flames. Non-reacting nitrogen is used as the diluent gas in the fuel stream. Unitary Lewis number provides a common thermal and mass diffusivity from which to create scalar dissipation rate. Both stable and unstable branches of flammability curves (S-curves) are calculated with three vorticity levels and plotted against multiple input and output parameters. The description of the three-dimensional flamelet structure, allowing vorticity and variable density to produce a centrifugal effect, is seen to be necessary for an accurate determination of the $\mathrm{H_2O}$ production rate when ambient inflow strain rate $(S^*)$ and vorticity $(\omega)$ are chosen as the key parameters. Maximum temperature and integrated $\mathrm{H_2O}$ production rate each nearly collapse to a single curve when plotted versus maximum scalar dissipation rate $(\chi_{max})$ but do not collapse when plotted versus the local maximum strain rate $(S^*_{local})$ or $S^*$. Additionally, $S^*_{local}$ and scalar dissipation rate $(\chi)$ depend strongly on vorticity and ambient inflow strain rate. It is argued that the controlling inputs for a flamelet embedded in a turbulent eddy are the ambient vorticity and strain rate which are thus the natural choice of parameterizing variables. These ambient quantities can be readily linked to the averaged or filtered turbulent flow by leveraging cascade theory, as opposed to local strain rate or scalar dissipation rate within the flame zone, which do not have a widely accepted, first-principles scaling connection to the turbulence cascade.
\end{abstract}

\end{frontmatter}
\newpage
\textbf{\textit{Novelty and Significance Statement}} \\
A novel, three-dimensional, rotational flamelet model is formulated with certain constraints from prior models but some new physics. Specifically, multi-step chemical kinetic models are added and it is shown that including vorticity and three-dimensional strain rates changes the ambient and local strain rates at which diffusion flames extinguish. The determination of scalar dissipation rate depends on vorticity and applied strain rate; however, flammability curves plotted versus maximum scalar dissipation rate collapse. This collapse is of limited value from the perspective of the authors because there is no known first principles scaling law for scalar dissipation rate to couple it to the resolved scale. Furthermore, the collapse does not justify assuming a functional form for scalar dissipation rate in place of solving the momentum equations with vorticity. This is significant because it provides a basis for which LES and RANS simulations may more accurately predict combustion dynamics via a coupling based on well-established scaling laws.

\textbf{\textit{Author Contributions}} \\
WH - Performed research, analyzed data, wrote the paper\\
XS - Analyzed data, edited the paper\\
WS - Analyzed data, edited the paper\\

\ifdefined \wordcount
\clearpage
\fi
\newpage
\section{Introduction\label{sec:introduction}} \addvspace{10pt}
Combustion occurring in practical engineering flows is subject to turbulence. In such flows, chemical reactions predominantly occur on the smallest turbulent length scales, within an order of magnitude of the Kolmogorov scale. It is well known that magnitudes of velocity gradients increase with decreasing length and velocity scales because the Kolmogorov length scale, with Reynolds number (Re), behaves as $~Re^{-3/4}$, and decays faster than the velocity scale, which behaves as $~Re^{-1/4}$. Therefore, the various velocity derivatives, three principal-normal-strain rates and vorticity, impact local flame structure, extinction, and re-ignition. The authors believe the accuracy of RANS and LES simulations may be improved with new methods that include these velocity derivatives in flamelet models and their coupling procedures.

Past models of such flames are based on the flamelet concept pioneered by Williams \cite{Williams1975} which considers turbulent flames existing in the layer between oxidizer and fuel interfaces as an ensemble of thin, highly-sheared, one dimensional, diffusive-reactive zones. The flamelet approximation is governed by the assumption that chemical and diffusive time scales are sufficiently short, such that a quasi-steady assumption can be made for each flamelet. We treat flamelets subject to the largest strain rate experienced in the flowfield; therefore, the relevant length scale is the Kolmogorov scale. If the reaction rate in a given area were too low compared to the diffusion rate (which is a function of the strain rate), a diffusion flame could only occur on a larger scale. This larger scale would not be laminar since smaller turbulent eddies would exist within it. Therefore, a Kolmogorov-scale model is the only laminar model relevant to turbulent flows. Assuming diffusivity and kinematic viscosity are of the same order, the mixing layer length will be on the same order of magnitude as the Kolmogorov scale. We consider large Damkohler number (reaction rate $>$ strain rate), so that a diffusion flame can occur. Thus, our reaction zone thickness is smaller than the diffusion length scale. Of course, premixed flames in turbulent flows can be larger than the smallest eddies \cite{Dunn1,Dunn2}, subjecting them to turbulent straining. Premixed flames of this type are beyond the scope of this work but have been addressed in the literature, generally under the classification of ``flamelet generated manifolds'' (FGM) \cite{PoinsotVeynanteCandel, Colin, Oijen1,Oijen2,Ketelheun}. Another method existing for premixed flames is ``flame prolongation of intrinsic low-dimensional manifold'' (FPI) \cite{Gicquel}. Both FGM and FPI methods are based on un-strained premixed flames. Some attempts have been made to extend FGM and FPI to diffusion and partially premixed flames with promising results \cite{Fiorina}. Other non-premixed flame models that have no intrinsic application to turbulence exist, such as the ChemKin OPPDIFF and Cantera difflame codes which solve a counterflow diffusion flame issuing from opposing nozzles \cite{CHEMKIN,Cantera}. These models are two-dimensional planar or axisymmetric and do not account for strain rates normal to the nozzle axes (transverse strain rates) nor vorticity. 

Flamelet models frequently used in RANS and LES simulations commonly make two-dimensional planar or axisymmetric assumptions, allowing the governing system of equations to be reduced to one-dimensional ordinary differential equations (ODEs). Vorticity and transverse strain rates also are not included or discussed. A transformation to mixture fraction space is commonly made to simplify the computation by removing the chemical source term. In the progress variable approach, upon which many recent flamelet models are based, Pierce and Moin \cite{PierceThesis,Pierce_Moin} acknowledge that Peters' practice \cite{Peters1984,Peters2000} of prescribing counterflow transverse velocity via a fixed strain rate, $u(x) = -S_1^*x$, is not valid across the entire domain as it violates the continuity equation, ignoring important dilatation effects. Pierce and Moin adopt zero advection in its place, and apply Dirichlet boundary conditions to species and enthalpy at a finite separation distance, $y = 0$ and $y = L$ in lieu of solving the momentum equations. This simulates counterflow streams by capping the diffusive domain, creating scalar gradients from which scalar dissipation rate can be defined. The scalar dissipation rate obtained in this manner is then strongly dependent on the choice of boundary conditions. It is then stated, implicitly or explicitly in models derived from the progress variable approach, that the local flame structure is parameterized by scalar dissipation rate (SDR) and the progress variable. The flamelet solutions are coupled to the resolved scale using the presumed PDF approach. A weighted sum of mass fractions is chosen as the progress variable, from which the flamelet outputs, parameterized by SDR, are convoluted with the presumed PDF, which itself, is parameterized by SDR, progress variable, and mixture fraction. Thus, a correlation, involving assumptions, is made between the mixture fraction and value of weighted mass fractions on the resolved scale, and the stoichiometric SDR in the flame zone of the flamelet. This may explain why RANS/LES sometimes fails to match experiments and DNS. 

Wang et al. performed DNS studies of turbulent spray flames to evaluate the applicability of various scaling laws for scalar dissipation rate and its probability density function \cite{WangKronenbergTufanoStein2017,WangKronenbergStein2019}. They investigated both droplet arrays and free droplets dependent on turbulent convection and, for the latter in particular, found that scaling laws involving corrected gaussian shape functions and $\beta$-PDFs for Kolmogorov-scale SDR fail to correctly emulate DNS data. They propose two additional scaling laws having better agreement. Ameen et al. \cite{Ameen2015} conducted an evaluation between simulations using a gaussian profile for SDR and DNS solutions for SDR. While focused on the change in SDR due to autoignition, they draw a similar conclusion that a gaussian assumption for SDR does not represent the physics well. While we are not dealing with two-phase mixtures or autoignition in the present work, this literature suggests that the universal gaussian assumption for SDR as a function of mixture fraction is inadequate. To address this, we hypothesize that scaling resolved flow velocity derivatives to ambient values relative to the flame, and then determining flame structure via sub-scale physics subject to these ambient boundary conditions, including vorticity and three-dimensional strain rates, may better mimic the physical process, ultimately improving simulation accuracy. This is our principal motivation for studying the effects of vorticity on flamelets. However, the potential benefits of this model in RANS and LES are currently undetermined and will be addressed in future work.

Many important improvements to non-premixed flamelet theory have been made in the last two decades. The original theory, now over 40 years old: 1) was limited to 2D planar and axisymmetric configurations; 2) assumed constant density in the momentum equation and thus a linear velocity variation through the flamelet rendering the momentum equation moot; 3) did not include vorticity or transverse strain rates since the momentum equation was not solved; 4) was limited to a unitary Lewis number (Fickian) diffusion process such that the mixture fraction was monotonic; 5) neglected curvature effects. Complete solution of the momentum equation with proper consideration of vorticity was not addressed prior to Sirignano \cite{Sirignano1}, however, many of these other restrictions have been removed. The assumption of a 1D axisymmetric or 2D planar configuration was removed through a transformation from 3D physical space to a 3D mixture fraction space ($x_1,x_2,x_3 \rightarrow Z_1,Z_2,Z_3$). First proposed by Peters \cite{Peters1984}, the theory was later improved by Xuan et al. \cite{XuanBlanquartMueller2014} taking tangential transport processes and curvature into account. The assumption of constant density in the momentum equation was relaxed by Kim and Williams \cite{KimWilliams1993} who obtain an approximate expression for scalar dissipation rate considering how heat release and composition produce density variation through the flame. However, Kim and Williams make other assumptions, e.g., $\rho h$ and $\rho D$ are constant and $L_e = 1$, among others which they acknowledge also need to be relaxed. Among the most critical and restrictive assumptions that have since been removed is the assumption of unitary Lewis number. With $L_e = 1$, the effects of both differential and preferential diffusion are negated. While an acceptable approximation in certain cases, this leads to unacceptable discrepancies from detailed transport models in many practical engineering situations, particularly those with highly diffusive fuels or a combination of species with a large variance in molecular weight \cite{Jiang2023,DONINI,Jiang2015,WuIhmeZhao2021,Gupta2014,Gierth2018}. Pitsch and Peters \cite{PitschPeters1998} introduced differential diffusion effects via a re-definition of mixture fraction whereby it no longer is a weighted sum of reactive scalars. Instead, they define mixture fraction via a conservation equation and state that taking the mixture fraction diffusivity equal to the thermal diffusivity produces a sufficiently accurate flame structure; however, it has been shown that this model over-predicts differential diffusion effects in turbulent jet flames since it does not factor in Reynolds number \cite{Pitsch2000,Wang2016}. The equations they obtain via a transformation from physical space to mixture fraction space are exact with respect to the pre-transformed equations if scalar dissipation rate as a function of mixture fraction, i.e., SDR(Z), is known. Herein lies the problem; one cannot know SDR(Z) without solving the momentum equations with vorticity, variable density, nor any other simplifications; a point clearly stated in Pitsch and Peters' conclusion ``The transformation has been shown to be exact, if the scalar dissipation rate is calculated as a function of the mixture fraction from the full set of fluid dynamic equations'' \cite{PitschPeters1998}. Other authors have also addressed differential diffusion effects. Wang \cite{Wang2016} identifies the influence of Reynolds number on differential diffusion showing analytically that molecular diffusion is dominant at low Reynolds numbers while turbulent diffusion is dominant at high Reynolds numbers. Consequently, he introduces the ratio of molecular to turbulent diffusion as an explicit parameter which has been shown to better approximate flame structure as compared to unity Lewis number and variable Lewis number models; however, results still showed deviation from experiments \cite{Wang2016,Gierth2018}. Following the work of Wang and Gierth et al., Jiang et al. \cite{Jiang2023} proposed the species-weighted flamelet (SWF) model, based on the FPV model, which, while similar to \cite{Wang2016,Gierth2018}, uses detailed molecular diffusion in addition to the Lewis number dependence on Reynolds number. The SWF model includes unity Lewis number (ULF) ($R_e \rightarrow \infty$) and variable Lewis number (VLF) ($R_e \rightarrow 0$) databases which are combined in a weighted average to approximate the effects of differential molecular diffusion. They found that differential diffusion notably alters profiles of temperature and species with LES results differing from experimental results \cite{Sevault2012} by less than 6$\%$. The SWF model better captures the inter-related physics of combustion intensity, turbulence intensity, mixture fraction, and differential diffusion.

Curvature effects, which were not present in classical theories \cite{Peters1984,Peters2000,PitschPeters1998,Pierce_Moin}, have been added to flamelet models over the last two decades. As the name suggests, curvature effects describe the implications of a curved flame front as opposed to the flat flame fronts originally considered. In a turbulent combustion domain, it is very conceivable that a flame could be distorted by an eddy, depending on the size of that eddy, and that the mathematical framework based on a flat flame would then breakdown. Xuan et al., discussed previously, derived a mathematical framework for three-dimensional (in mixture fraction space) flamelets considering curvature and differential diffusion \cite{XuanBlanquartMueller2014}. They found that curvature effects must be considered ``when the radius of curvature is comparable or smaller than the local flame thickness'' \cite{XuanBlanquartMueller2014} where flame thickness is defined as $l_F = (\Delta Z)_F/|\nabla Z|_{st}$. This is an important addition to flamelet theory and one that helps steer our analysis. We conclude that curvature effects may be neglected in the present model on the basis that no fluid structures with length smaller than the Kolmogorov scale exist in the flow.

Xuan et al. also found that effects of curvature do not impact transport physics if unity Lewis number is assumed but do impact the physics if differential diffusion effects are considered. We do not limit our model to constant or unitary Lewis number but can impose such conditions if desired. A brief note on curvature and vorticity is also important. The curvature effects in flamelet models, discussed in recent years, are fundamentally different from the vorticity effect we incorporate. Curvature effects describe distortion of the mixing layer and flame which produce scalar transport in directions tangential to the counterflow axis. They do not include a centrifugal term in the momentum equations as we have done. While flow vorticity can contribute to the degree of curvature of a scalar iso-surface in turbulence, the inclusion of curvature terms in the flamelet equations is not a substitute for the solution of the momentum equation with vorticity. A model adapted from the present, including curvature effects, is certainly an interest for future study.

The FPV model and its derivatives have been great improvements to simulations of turbulent reacting flows, better predicting unsteady effects such as quenching, re-ignition, and flame liftoff, while providing a relatively simple scale coupling procedure that does not add excessive computational challenge to the simulations. A portion of the realized improvements of the FPV model stem from the inclusion of both the stable and unstable branches of the S-shaped curve. This allows the model to better capture unsteady effects and we seek to retain this feature, although, flamelet models can and have been developed using only the stable branch. However, alternate approaches and improvements are still allowed, some of which we seek to address. Furthermore, the mixture fraction and SDR are only useful concepts with specific and sometimes limiting assumptions about the diffusion process. Specifically, our aim is to solve the three-dimensional momentum equations for the flamelet with vortex stretching to obtain the velocity field and parameterize the subsequent flamelet solutions via velocity gradients that have established scaling laws such that the progress variable and its correlation with stoichiometric SDR are not needed. Unless the momentum equations give the correct velocity field, one cannot expect to obtain the correct SDR. We show in this manuscript that vortex stretching is crucial in the solution of the momentum equations when strain rate and vorticity are directly applied to the model and that different global inputs i.e., ambient inflow strain rate and vorticity, can produce the same scalar dissipation rate. 

The present model builds upon two presuppositions regarding the nature of the flow across the length scales. First, there exists, in the literature on turbulence, established scaling laws for velocity and velocity derivatives, making these the ``first principles'' choices for coupling input parameters to the flamelet model. Second, we hypothesize that the flamelet scale solution ``sees'' these as far-field or ambient conditions such that a three-dimensional picture of velocity, strain rate, and vorticity exist as boundary conditions for the counterflow. Thus, we aim to couple the flamelet model to the resolved scales via established scaling laws for velocity derivatives without adding additional PDEs, such as that for progress variable, to the resolved scale. This preference is based on the abundance of evidence in the literature, both from experiment and direct numerical simulation that, for turbulent reacting and nonreacting flows, averages of normal strain rates, shear strain rates, and vorticity are important \cite{Nomura1,Nomura2,Boratav1996,Boratav1998,Ashurst}, but are not directly included in flamelet modeling.

Sirignano \cite{Sirignano1,Sirignano2} using one-step chemical kinetics and Fickian diffusion, showed that inclusion of vorticity causes a centrifugal effect that increases the flammability limit and reduces the mass burning rate. When three principal strain rates are included, i.e. neither a planar nor axisymmetric flow, the centrifugal effect is modified by the far-field strain rates' impact on mass efflux.

In this study, for the limited cases where mixture fraction and SDR are meaningful variables, we make four primary points: First, the determination of both SDR and local strain rate from first principles depends on both vorticity and applied strain rate. Flammability curves (S-curves) plotted versus SDR collapse as shown in Fig. \ref{fig3}; however, there is no known first-principles approach to relate SDR to resolved-scale behavior. Second, vorticity increases the maximum ambient strain rate above which a diffusion flame will extinguish. Third, the ambient transverse strain rates (those normal to the inflow axis) also influence the extinction strain rate. Lastly, we argue that scaling laws for velocity gradients are both more substantiated and more representative of the physics than procedures to relate SDR to the resolved scale; however, this is our opinion and not affirmed or disputed by our results or any existing literature.

\section{Methods\label{sec:sections}} \addvspace{10pt}
\subsection{Classical Flamelet Equations\label{subsec:subsection1}} \addvspace{10pt}
The classical flamelet equations for non-premixed combustion as presented by Peters \cite{Peters1984,Peters2000} make the simplification of Fickian diffusion which allows the energy and species continuity equations to both take the form of Eq. (1). With the diffusivities of all species equal, counter-gradient diffusion does not exist and the mixture fraction varies monotonically between the oxidizer and fuel streams. This is a very elegant and convenient assumption, allowing the governing equations of reactive scalars to be greatly simplified, taking the form of Eq. (2) where scalar dissipation rate, $\chi$, is the only critical coefficient of the equation, i.e., the term involving the Lewis number is removed.
\begin{equation}
    \rho \frac{\partial \psi_i}{\partial t} +\frac{1}{4}\left(1-\frac{1}{Le_i}\right)\rho\frac{\partial\chi}{\partial Z}\frac{\partial\psi_i}{\partial Z}= \frac{\rho \chi}{2}\frac{\partial^2 \psi_i}{\partial Z^2} + \dot{\omega}_i
\end{equation}
\begin{equation}
    \rho \frac{\partial \psi_i}{\partial t} = \frac{\rho \chi}{2}\frac{\partial^2 \psi_i}{\partial Z^2} + \dot{\omega}_i
\end{equation}
The dependent variables are $\psi = Y_1, Y_2,..., Y_N, T$ and a different source term ($\dot{\omega}_i$) exists for each equation. The mixture fraction is $Z$ and SDR is defined as $\chi = 2D|\partial Z/\partial x_j|^2$. This formulation is very elegant in its simplicity and has served the turbulent combustion community for decades. However, the formulation is limited by the simplified mass diffusion scheme. In situations where differential diffusion and counter-gradient diffusion are important, both the concepts of mixture fraction and scalar dissipation rate have significantly reduced utility and do not lead to a simple flamelet formulation. Furthermore, the dependence on SDR requires the creation of a contrived variable (the progress variable) and a presumed PDF to have a unique mapping technique between the resolved scale and the flamelet table.

\subsection{Governing Equations\label{subsec:subsection2}} 
\begin{figure}[h!]
\centering
\includegraphics[width=470pt]{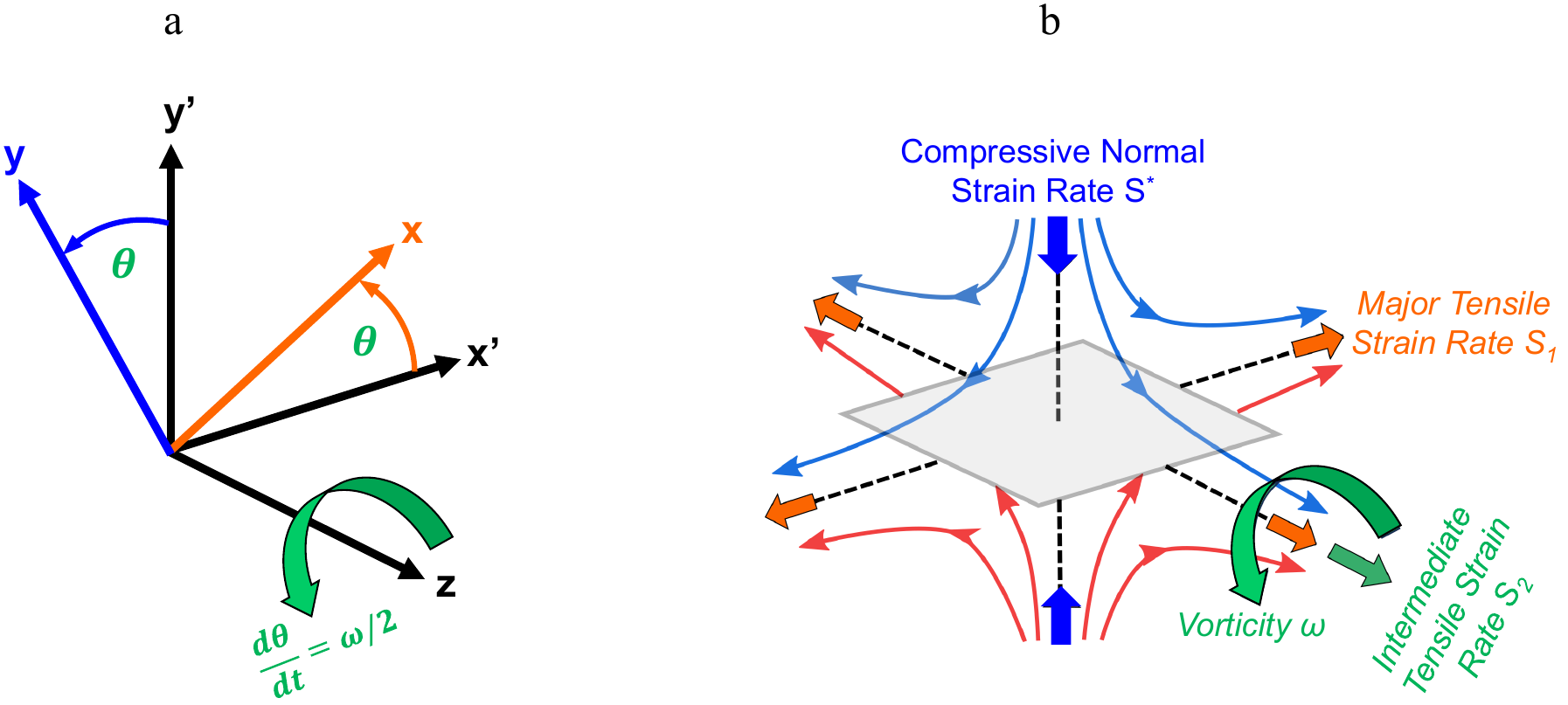}
\caption{\footnotesize Non-Newtonian transformation (\textbf{a}) and relative alignment of strain rates and vorticity with oxidizer inflow (light blue) and fuel inflow (red) (\textbf{b}).}
\label{fig1}
\end{figure}

\addvspace{10pt}
To incorporate vorticity $(\omega)$, a non-Newtonian reference frame is chosen that rotates about the vorticity vector at a rate of $d\theta/dt = \omega/2$ according to Fig. \ref{fig1}\textbf{a}. The governing equations for three-dimensional steady flow in this reference frame are listed below, Eq. (3-7) where $a_i = \langle x\omega^2/4,y\omega^2/4,0 \rangle$ is the centrifugal acceleration in the non-Newtonian reference frame. The low Mach number approximation is made to remove terms involving kinetic energy and viscous dissipation. The vorticity vector is chosen to align with the $z$ axis according to Fig. \ref{fig1}. Major compressive strain rate is aligned with the $y$ axis while major tensile strain is aligned with the $x$ axis. Intermediate tensile strain rate is aligned with the $z$ axis and vorticity is also aligned with the $z$ axis. This choice was made due to DNS findings \cite{Nomura1, Nomura2} that indicate that vorticity is statistically likely to align with the direction of intermediate tensile strain in non-premixed flames and that the direction of scalar gradients is normal to this direction, aligning with the direction of compressive strain. It is also noted that vorticity alignment with the intermediate tensile strain direction is more probable in the case of sheared turbulence rather than isotropic turbulence \cite{Nomura1}. Regardless of the configuration, two strain rates are likely to be extensional while only one is compressive.

\begin{equation}
    \frac{\partial (\rho u_j)}{\partial x_j} = 0
\end{equation}
\begin{equation}
    \rho u_j\frac{\partial u_i}{\partial x_j} + \frac{\partial p}{\partial x_i} = \frac{\partial \tau_{ij}}{\partial x_j} + \rho a_i
\end{equation}
\begin{equation}
    \tau_{ij} = \mu\left[\frac{\partial u_i}{\partial x_j} + \frac{\partial u_j}{\partial x_i} - \frac{2}{3}\delta_{ij}\frac{\partial u_k}{\partial x_k}\right]
\end{equation}
\begin{equation}
    \frac{\partial \left(\rho u_j h\right)}{\partial x_j} + \frac{\partial}{\partial x_j}\sum_{m=1}^N \rho v_m^d Y_m h_m = \frac{\partial}{\partial x_j}\left(\lambda\frac{\partial T}{\partial x_j}\right) - \rho\sum_{m=1}^N h_{f,m}\dot{\omega}_m
\end{equation}
\begin{equation}
    \frac{\partial \left(\rho u_j Y_m\right)}{\partial x_j} + \frac{\partial}{\partial x_j}\left(\rho Y_m v_m^d \right) = \rho \dot{\omega}_m
\end{equation}

In order to define the scalar dissipation rate commonly used in the literature, see Eq. (24), thermal and mass diffusivities for all species are taken to be equal. We assume unitary Lewis number which specifies the diffusivity at each point in the domain according to Eq. (8). Diffusion velocity and diffusion flux are defined according to Eq. (9-10). Using Taylor series approximations in classical fashion for transverse velocity components in a counterflow, i.e., Eq. (11-12), we obtain a similar solution described by ordinary differential equations, Eq. (15-19). The full derivation is given in \cite{Sirignano1,Sirignano2,Hellwig2024}.
\begin{equation}
D = \frac{\lambda}{\rho c_p}
\end{equation}
\begin{equation}
v_m^d = -\frac{1}{Y_m}D\frac{d Y_m}{d y}
\end{equation}
\begin{equation}
J_m = \rho Y_m v_m^d
\end{equation}
The similarity solution simplifies the solution to Eq. (3-7), reducing the scalar fields to functions of only the axial coordinate $y$. In doing so, we do not account for curvature effects. The velocities in the transverse directions, $x$ and $z$ are assumed to be functions of their respective spatial variables and of $y$ according to Eq. (11-12).
\begin{equation}
    u_x = S_1 x f_{1}^{'}
\end{equation}
\begin{equation}
    u_z = S_2 z f_{2}^{'}
\end{equation}
The two constants $S_1$ and $S_2$ are the far-field strain rates at $x = +\infty$ and $z = +\infty$, respectively, non-dimensionalized by the ambient strain rate at $y = +\infty$, $S^*$. For the remainder of this manuscript, $^*$ indicates a dimensional quantity. All other quantities are non-dimensional. Variables $f_{1}^{'}$ and $f_{2}^{'}$ are two of three similarity variables, all of which are functions of $y$ only. The third similarity variable is defined according to Eq. (13) from which the $y$ velocity component is found via Eq. (14).
\begin{equation}
    f = S_1 f_{1} + S_2 f_{2}
\end{equation}
\begin{equation}
    u_y = -\frac{f}{\rho}
\end{equation}
Employing a Howarth–Dorodnitsyn transformation from $y\,\to\,\eta$ i.e., $\eta \equiv$ \(\int_{0}^{y} \rho(y^{}) \,dy^{}\), allows the equations to be further simplified. After the transformation to $\eta$-space is made, we denote first derivatives with respect to $\eta$ using superscript $'$. Derivatives of higher order have additional ``primes'', e.g., a second derivative has superscript $''$. For a complete derivation of the similarity solution, the reader is encouraged to see Sirignano \cite{Sirignano1} pg. A21-11-16. The similar form of the governing equations and their boundary conditions are presented below, along with the equation of state. Equations (15-16) are the momentum equations in the $x$ and $z$ directions. The $y$ momentum equation is not needed to obtain the velocity field but can be used to obtain the pressure gradient.
\begin{equation}
    \rho \mu f_1{'''} + f_1{''}(\rho \mu){'} + f f_1{''} + S_1\left(\frac{1}{\rho} - (f_1{'})^2\right) + \frac{\omega^{2}}{4S_1}\left(1 - \frac{1}{\rho}\right) = 0
\end{equation}

\begin{equation}
    \rho \mu f_2{'''} + f_2{''}(\rho \mu){'} + f f_2{''} + S_2\left(\frac{1}{\rho} - (f_2{'})^2\right) = 0
\end{equation}

\begin{equation}
   c_p f T^{'} + (\rho\lambda T^{'})^{'} - T^{'}\sum_{m=1}^N J_m c_{p_m} - \sum_{m=1}^N h^0_m\dot{\omega}_m = 0
\end{equation}

\begin{equation}
    f Y_m^{'} - J_m^{'} + \dot{\omega}_m = 0 \;\; ; \;\; m = 1,2,3.....N
\end{equation}
\begin{equation}
    P = \rho R_{sp} T
\end{equation}
The boundary conditions for Eq. (13-17) given by Eq.(18-21).
\begin{equation}
    f_1^{'}(\infty) = 1 \;\; ; \;\;  f_1^{'}(-\infty) = \sqrt{\frac{1}{\rho_{-\infty}} + \left(\frac{\omega_k}{2S_1}\right)^2\left(1 - \frac{1}{\rho_{-\infty}}\right)} \;\; ; \;\; f_1(0) = 0
\end{equation}
\begin{equation}
    f_2^{'}(\infty) = 1 \;\; ; \;\;  f_2^{'}(-\infty) = \frac{1}{\sqrt{\rho_{-\infty}}} \;\; ; \;\; f_2(0) = 0
\end{equation}

\begin{equation}
    Y_m(\infty) = Y_{m,\infty} \;\; ; \;\; Y_m(-\infty) = Y_{m,-\infty}
\end{equation}

\begin{equation}
    T(\infty) = 1 \;\; ; \;\;  T(-\infty) = T_{-\infty}
\end{equation}

Realize that enstrophy, $\omega^{*2} = \omega^* \cdot \omega^*$ is actually the relevant variable. The direction of rotation in the $x-y$ plane is not relevant. Analytical and experimental studies relate enstrophy to the square and products of strain rates, concluding that they are of the same order of magnitude \cite{Betchov,Raynal,Zhu_Antonia_1996}. As such, we choose to normalize vorticity by normal compressive strain rate $S^*$, keeping this nondimensional vorticity of $O(1)$; $0 \leq \omega = \omega^*/S^* \leq \sqrt{1.5}$.

One key difference between our flamelet configuration and former models is that our geometry is three-dimensional. Specifically, three normal strain rates are allowed to differ and vorticity can be applied. Furthermore, we do not use a nozzle-counterflow as many opposed-flow flame solvers do. At infinity, we take a potential counterflow which removes the imposition of zero velocity gradients at a fixed nozzle separation distance. Through the similarity solution, this three-dimensional geometry may be solved with ODEs.

\subsection{Scalar Dissipation Rate and Mixture Fraction\label{subsec:subsection3}} \addvspace{10pt}
Instead of strain rate, scalar dissipation rate (SDR), $\chi$, is commonly used to classify extinction in flamelet modeling. Beginning with Eq. (25), conservation of mixture fraction $Z$, an ODE for SDR can be derived in terms of the similarity variable $f$ and other quantities known once Eq. (15-19) have been solved. The resultant equation, Eq. (26), is a linear, homogeneous ODE with known boundary conditions when formulated in terms of the mixture fraction which varies from 0 to 1.
\begin{equation}
    \chi \equiv 2D\left(\frac{\partial Z}{\partial x_i}\right)^2 = 2D\left(\frac{dZ}{dy}\right)^2
\end{equation}
\begin{equation}
    \rho u_y \frac{dZ}{dy} = \frac{d}{dy}\left(\rho D \frac{dZ}{dy}\right)
\end{equation}
Multiplication by $2(dZ/dy)$ results in a first-order ODE for SDR.
\begin{equation}
    \left[ \frac{2f}{\rho D} + \frac{2}{\rho D}\frac{d(\rho D)}{dy} - \frac{1}{D} \frac{dD}{dy} \right] \chi + \frac{d \chi}{dy} = 0
\end{equation}
By defining $C$ as the constant of integration in Eq. (26) and integrating, Eq. (27) is found for $\chi$.
\begin{equation}
    \chi = \frac{C}{\rho^2 D} exp\left(-2\int_{0}^{\eta} \frac{f}{\rho^2 D} \,d \eta^{'}\right)
\end{equation}
Replacing $\chi$ in Eq. (27) with $2D \left(dZ/dy\right)^2$ through the definition of SDR and integrating from $-\infty\,\to\,\infty$ in $\eta$-space such that $Z$ varies from $0\,\to\,1$, allows $C$ to be determined.
\begin{equation}
    C = \frac{2}{\left(\int_{-\infty}^{\infty} \frac{1}{\rho^2 D} exp\left[-\int_{0}^{\eta} \frac{f}{\rho^2 D} \,d \eta^{'} \ \right] \,d \eta \ \right)^{2}}
\end{equation}
\subsection{Chemical Kinetics Model\label{subsec:subsection4}} \addvspace{10pt}
Multi-step chemical kinetics were incorporated through Cantera \cite{Cantera}. A skeletal version of the Foundational Fuel Chemistry Model 1.0 (ffcm-1) \cite{ffcm1} for hydrogen-oxygen combustion was used. This model also includes nitrogen as a non-reacting species for a total of 9 species and 25 reactions. The skeletal model was chosen to decrease computation time. 

\subsection{Numerical Methods\label{subsec:subsection5}} \addvspace{10pt}
The governing equations were solved using a hybrid pseudo-time stepping and Newton method approach implemented in Matlab. This code is a second-order finite-difference code utilizing the Cantera Matlab package to introduce multi-step chemical kinetics and integrated Matlab functions to solve the Matrix equations involved in the Newton method. 

\section{Results\label{sec:figtabeqn}} \addvspace{10pt}
\subsection{Equal Transverse Strain Rates ($S_1 = S_2$)\label{subsec:ResultsSec1}} \addvspace{10pt}
The counterflow diffusion flame studied in the following sections uses oxygen ($\mathrm{O_2}$) at 300 K issuing from $y = -\infty$ and an equimolar mixture of hydrogen ($\mathrm{H_2}$) and nitrogen ($\mathrm{N_2}$), also at 300 K, issuing from $y = +\infty$. The pressure is set at 10 atmospheres. Calculations were performed for three values of non-dimensional vorticity, $\omega = 0, 1, \sqrt{1.5}$, with $S_1 = S_2 = 1/2$ to simulate the existing flamelet models' assumption of axisymmetry. Note that our calculations are axisymmetric only when $S_1 = S_2 = 1/2$ and $\omega = 0$. When $\omega > 0$, vortex stretching implies non-symmetric behavior.

\begin{figure}[h!]
\centering
\includegraphics[width=250pt]{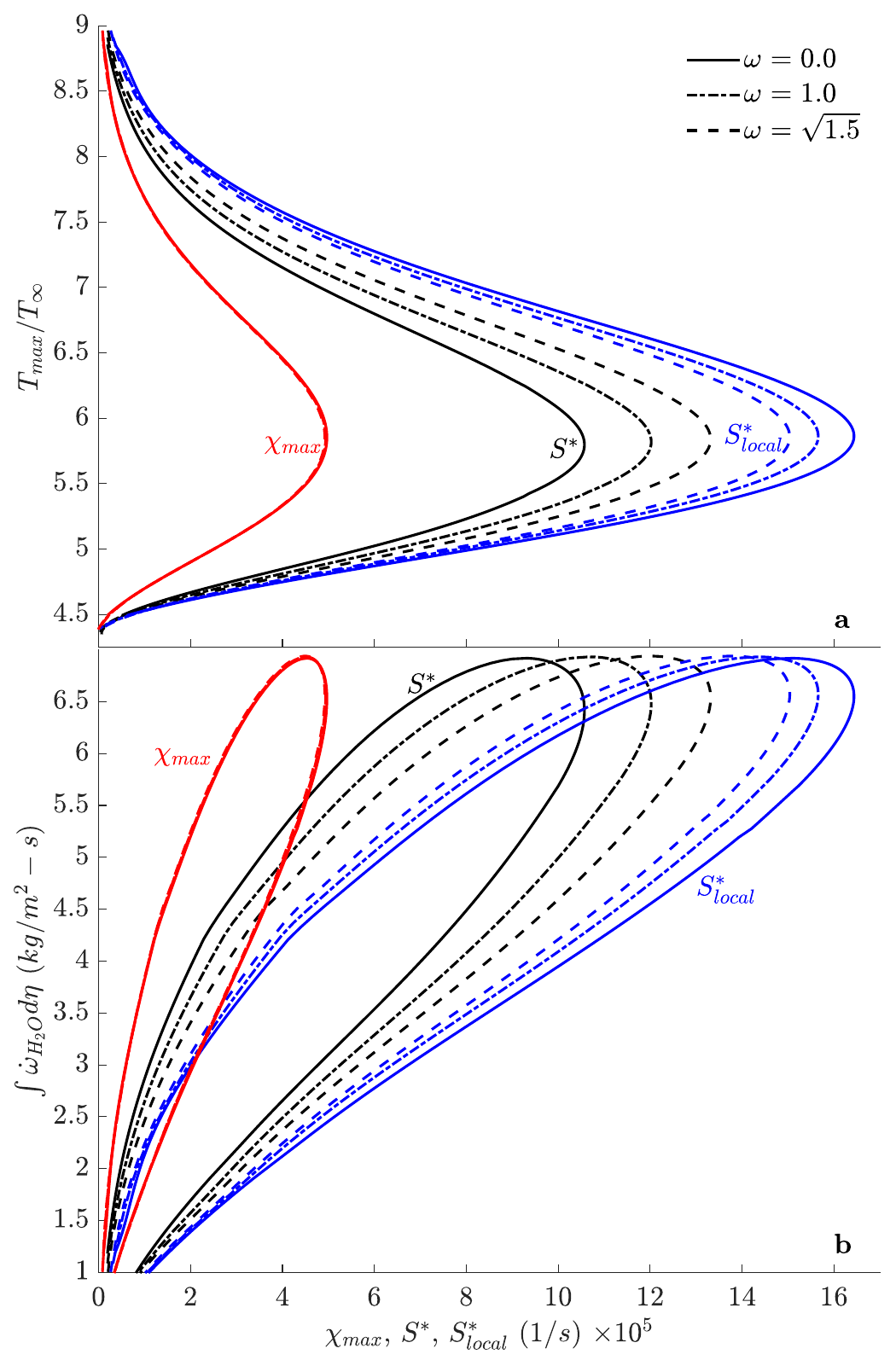}
\caption{\footnotesize Extinction curves versus: maximum scalar dissipation rate, $\chi_{max}$ (red); ambient strain rate, $S^*$ (black); maximum local strain rate, $S^{*}_{local}$ (blue).}
\label{fig2}
\end{figure}

First, extinction curves, or S-curves, were calculated to determine the flammability limits at each value of $\omega$. Figure~\ref{fig2}\textbf{a} shows these extinction curves plotted in three ways from left to right: against the maximum SDR, $\chi_{max}$, the ambient (i.e., inflow) axial strain rate at $y = +\infty$, $S^*$, and the maximum or local axial strain rate in the flame zone, $S^{*}_{local}$.

The black curves show a substantial impact of vorticity on the extinction ambient strain rate of the flames with a $26\%$ increase occurring between $\omega = 0 \rightarrow \sqrt{1.5}$. When local maximum strain rate is used on the abscissa (blue curves), those with $\omega \neq 0$ fall to the left of that with $\omega = 0$ because vorticity-induced changes to the velocity field decrease local strain rate with increasing vorticity. Since the temperature at the extinction point is nearly unchanged by vorticity, those curves with decreased local strain rate fall to the right with increasing vorticity when plotted against the inflow strain rate. When the S-curve is plotted versus maximum SDR, a near-total collapse occurs, although the order of curves follows those versus maximum local strain rate. Figure~\ref{fig2}\textbf{b} plots the integrated burning rate of water (\chemfig{H_2O}) along the S-curves in Fig.~\ref{fig2}\textbf{a}. Again, the solution is highly dependent on vorticity magnitude when plotted versus $S^*$ and $S^{*}_{local}$ and significantly collapses versus $\chi_{max}$. Actually, maximum temperature is a better predictor than SDR to capture ``collapse". From Fig.~\ref{fig2}, we see that one value of $\omega$ together with one value of $S^*$, $S^{*}_{local}$, or $\chi_{max}$ determines two values of $T_{max}$ and \chemfig{H_2O} production rate. However, one value of $T_{max}$ with one value of $\omega$ gives distinct values for $S^*$, $S^{*}_{local}$, $\chi_{max}$, and \chemfig{H_2O} production rate.
\begin{figure}[h!]
\centering
\includegraphics[width=250pt]{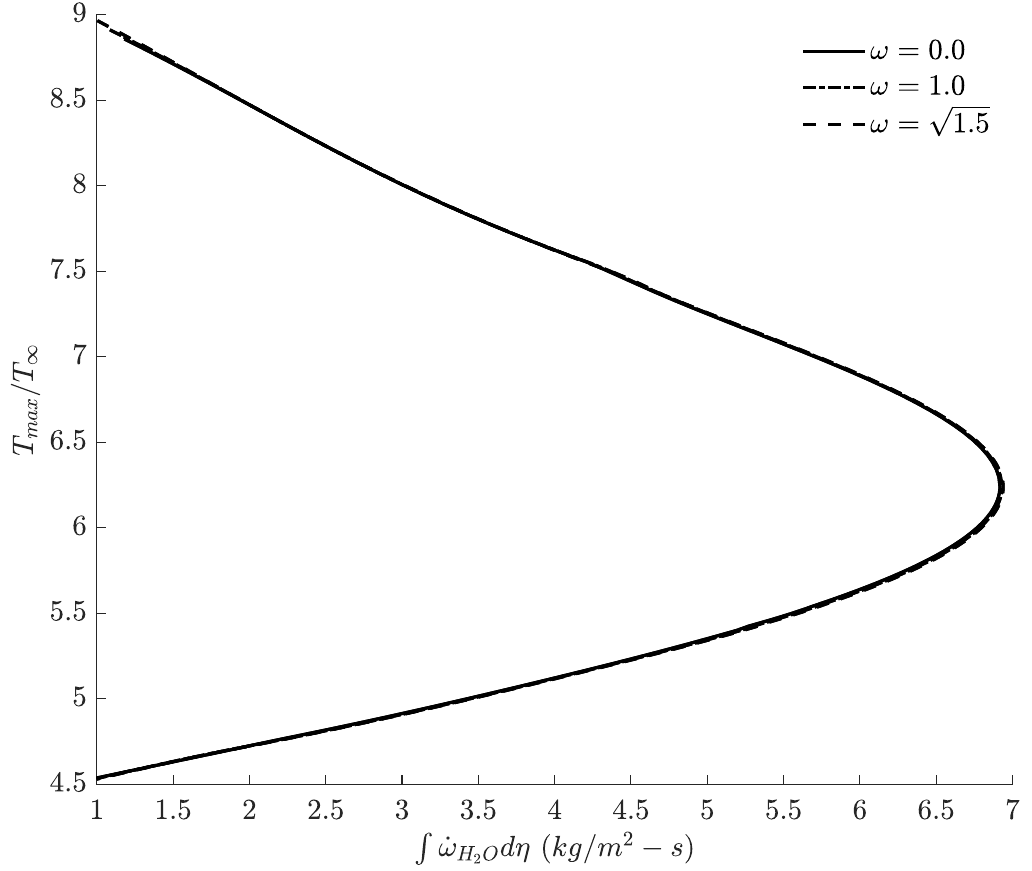}
\caption{\footnotesize Temperature versus integrated \protect\chemfig{H_2O} burning rate.}
\label{fig3}
\end{figure}
\begin{figure}[h!]
\centering
\includegraphics[width=250pt]{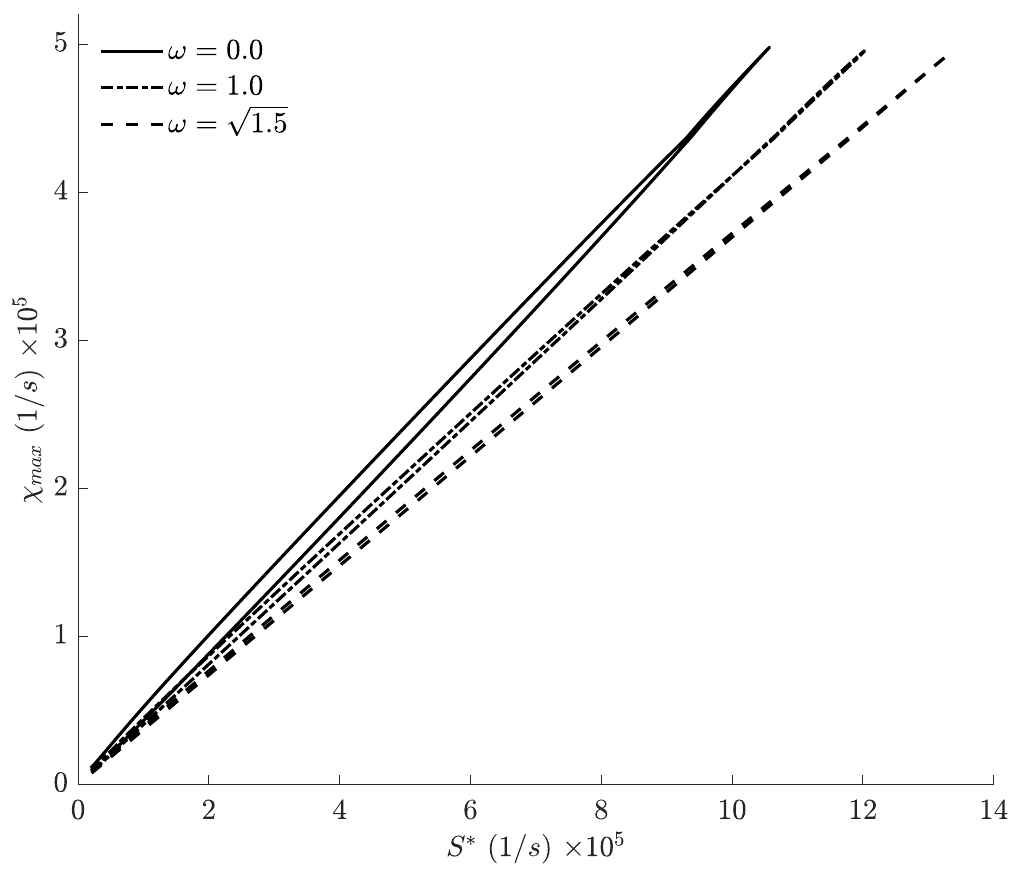}
\caption{\footnotesize SDR versus Ambient Strain Rate}
\label{fig4}
\end{figure}

Many flamelet approaches include only the stable branch of the S-curve. This is a viable tactic in some scenarios and simplifies computation by requiring only SDR or, in our case, strain rate and vorticity. If one wants to use the unstable branch as well, which more accurately captures extinction, reignition, pressure oscillation, and other unsteady effects \cite{NguyenSirignano2018}, an additional parameter is necessary to distinguish the branches, such as the progress variable or maximum temperature.

If output parameters are plotted versus output parameters, curves with varying vorticities collapse; see Fig.~\ref{fig3} and Fig.~\ref{fig2} (red). However, no output parameters plotted versus the two input parameters, $S^*$ and $\omega$, collapse; see Fig.~\ref{fig4} and Fig.~\ref{fig2} (black, blue). Obviously, integrated production rate versus maximum temperature would collapse if plotted. It is not surprising that one output plotted against another output collapses. This collapse results from the system having several input parameters, two of which ($S^*$ and $\omega$) similarly alter behavior. It is the reciprocal strain rate in the flame zone, i.e., residence time, that predominantly affects combustion intensity. All things being equal, an increase in $S^*$ increases flame zone strain rate, decreasing residence time while an increase in $\omega$ does the opposite. It is just as reasonable to create a flamelet theory using maximum temperature as the primary parameter as it is to make maximum SDR (or stoichiometric SDR) the primary parameter. Temperature, in fact, was used by Knudsen and Pitsch \cite{KnudsenPitsch2009} as the unique parameter is distinguishing between non-premixed flamelet solutions in the creation of a flame index for a multi-regime flamelet model. Maximum temperature, maximum SDR, and integrated production rate are all outputs dependent on vorticity and applied strain rate. The scaling of any of these output parameters given resolved-scale quantities is undetermined following first principles; rather, some contrivance such as a progress variable and an assumed functional form of $\chi(Z)$ is required. It is well known that scaling rules can be found for applied strain rate and vorticity, so, we hypothesize that utilizing these scaling rules may provide a more robust coupling procedure between sub-scale and resolved scale quantities.

The tendency of curves to collapse might at first appear convenient, allowing solutions to be parameterized by $\chi_{max}$ instead of $S^*$ and $\omega$, and indeed if only stable branch solutions are considered, and one is willing to bypass direct solution of the momentum and continuity equations by assuming some $\chi(Z)$, $\chi_{max}$ is adequate; however, some issues persist. First, as discussed in the introduction, recent flamelet approaches utilizing a transformation to 3D mixture fraction space are exact with respect to their pre-transformed equations, however, any flaw in the pre-transformed equations or in the assumption of variables in the post-transformed equations will cause a deviation from the real physics. In the flamelet form, similar to Eq. 1, the effects of the momentum and continuity equations are contained in scalar dissipation rate, $\chi$. That is okay provided the $\chi$ used to solve Eq. 1 came from physically complete conservation equations. The vast majority of existing models, even the most current, while having a correct ``flamelet equation'' in terms of $\chi$ then inject some sort of assumed gaussian variance for $\chi$ which completely ignores vorticity, and sometimes variable density, in the momentum equation. Secondly, at the fundamental level as a physically meaningful parameter, SDR precludes preferential (unequal Lewis numbers) and differential (non-unitary Lewis numbers) diffusion \cite{Moshe} since its definition stipulates a common diffusivity. Incorporating detailed transport allows species and heat to locally re-distribute which notably impacts laminar flames \cite{Jiang2023,DONINI,Jiang2015,WuIhmeZhao2021}. Corrections to models assuming constant Lewis number can be made to better capture the tendencies of differential diffusion \cite{Jiang2023,Gupta2014,PitschPeters1998,Gierth2018,Wang2016}, but it becomes difficult to create a useful and clearly understood conserved scalar. An alternate approach, governing a conserved scalar with a transport equation in which the diffusivity is equal to the thermal diffusivity but differs from the species diffusivity, was introduced by Pitsch and Peters \cite{PitschPeters1998}. Although it is named the ``mixture fraction," it differs from the classical definition and has diminished physical meaning. A complicated set of differential equations are yielded, resulting in a scalar dissipation rate with unclear physical interpretation and thus the problem of coupling it to the resolved scale via a ``first principles'' method persists. These improvements to existing flamelet models clearly show the importance of differential diffusion. Present flamelet theory, including differential and preferential diffusion, for example Jiang et al. \cite{Jiang2023}, are definite improvements on the classical theory. Therefore, differential and preferential diffusion alone, are not the major improvements pursued here. Rather it is the accruement of underlying assumptions, not all of which have been relaxed, which inspired our rotational flamelet model. It is not based on existing models and is constructed from the ground up with no constraints on Lewis number, geometry, velocity derivatives, or thermo-physical relations. We hypothesize that the rotational flamelet model will yield an improvement in the context of RANS and LES because it addresses previously excluded physics.

Furthermore, in application of a flamelet model to turbulent combustion, the cascades of length, velocity, and velocity derivatives (strain rate and vorticity) are much better understood than the cascades of scalar gradients \cite{Johnson2024}. Therefore, we believe it is worth investigating whether parameterizing flamelet solutions with strain rate and vorticity may improve accuracy in RANS and LES. It is noteworthy, however, that employing strain rate and vorticity adds an additional matrix parameter and is thus more computationally expensive. Also note that, like SDR, strain rate and vorticity cannot distinguish the stable from the unstable branch.
\begin{figure}[h!]
\centering
\includegraphics[width=250pt]{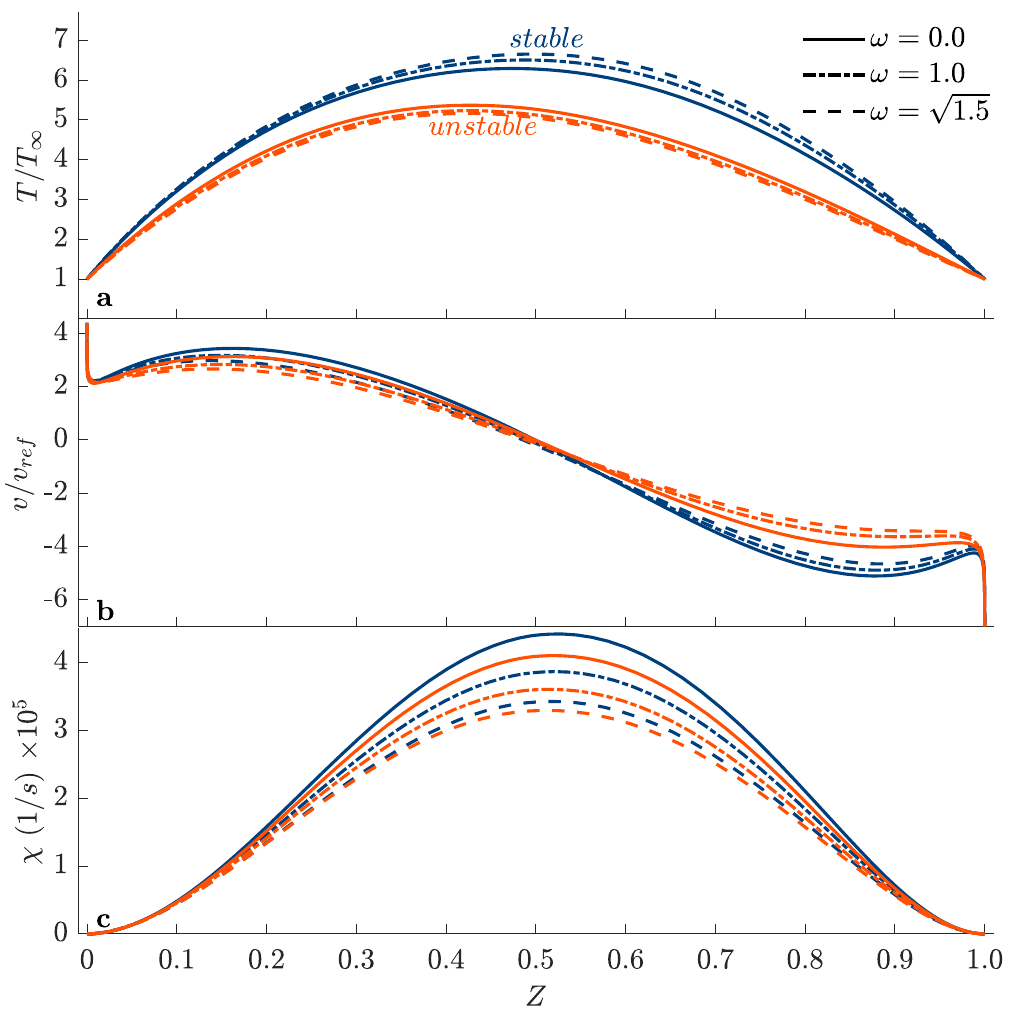}
\caption{\footnotesize Temperature and axial velocity versus mixture fraction. Stable branch (navy); unstable branch (orange). $S^* = 903,500$ $1/s$.}
\label{fig5}
\end{figure}

The importance of solving the momentum equations with vorticity is illustrated in Fig.~\ref{fig5}. Although the velocity curves have the same ambient strain rate, $S^*$ = 903,500 $1/s$, their slopes and magnitudes differ as a result of mass efflux changes caused by the centrifugal action of vorticity coupled with the density gradients throughout the counterflow. With vorticity, the lighter fluid (products of combustion) tends to flow outward aligned with the vorticity axis as this direction does not experience centrifugal force. The denser fluid is thrown outward in the plane normal to the vorticity axis, counteracting the incoming pressure gradient. This results in a lower strain rate at the stagnation point and thus increased residence time, which is responsible for increased temperatures on the stable branch. Opposite to temperature, SDR decreases as vorticity increases because the lower velocity gradients cause lower mixture fraction gradients. This result, shown in Fig.~\ref{fig5}\textbf{c}, clearly indicates that vorticity must be considered to accurately determine SDR. The existing progress variable method relies on taking scalar values at the point of stoichiometric mixture fraction to define the progress variable and parameterize the S-curve \cite{Pierce_Moin}. These scalars are typically weighted mass fractions and SDR. We show here that such values depend on $\omega$ and $S^*$ and may therefore benefit by not being characterized by one progress variable. In the case that all mass diffusivities are equal to each other but not equal to thermal diffusivity, i.e., $L_e\neq 1$, another term appears in the scalar transport equations and thus an additional parameter is needed to classify solutions. Even if the assumptions involved in deriving Eq. (1) were inconsequential (which they are not), the issue of coupling the flamelet solutions to the resolved flow via established scaling laws remains.

\begin{figure*}[h!]
\centering
\includegraphics[width=408pt]{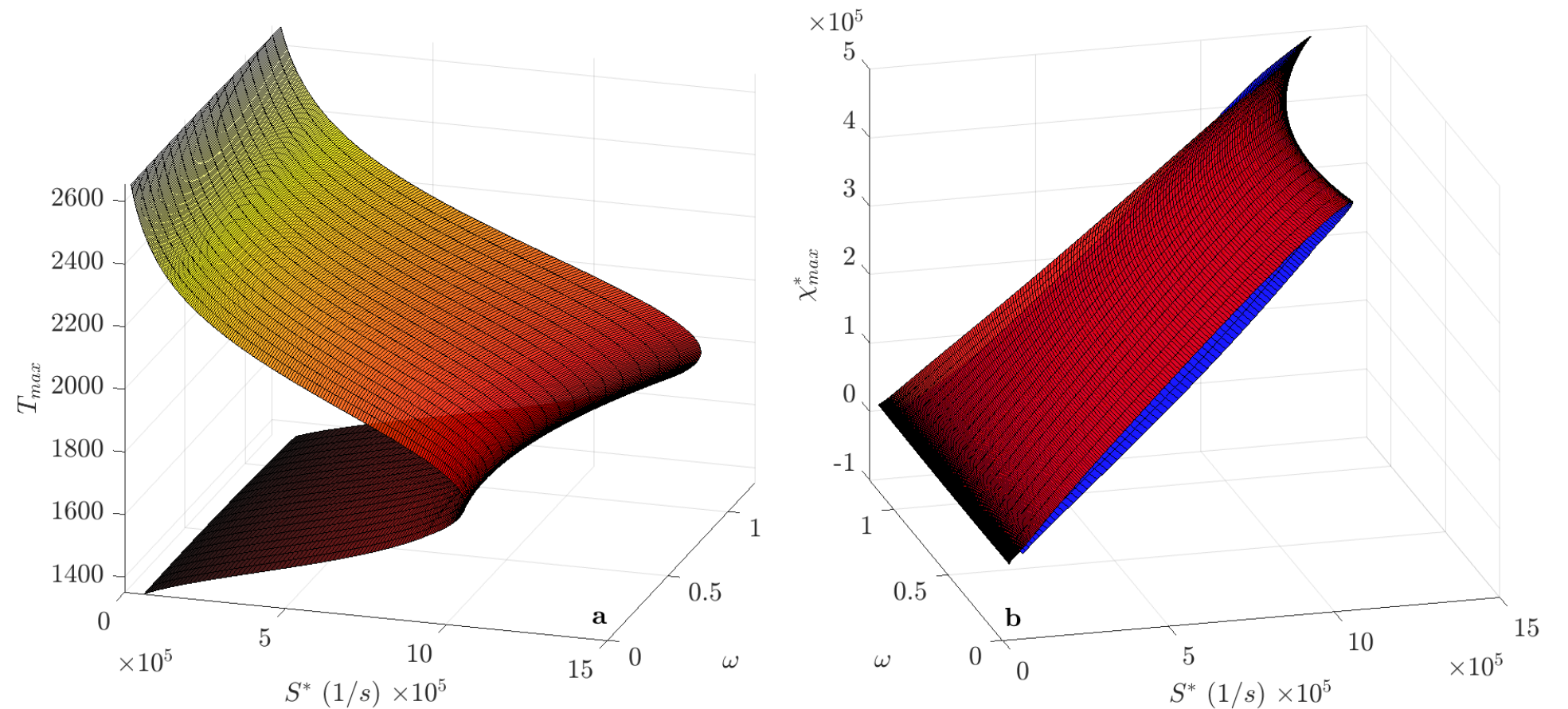}
\vspace{10 pt}
\caption{\footnotesize Three-dimensional extinction surface (\textbf{a}) and SDR surface (\textbf{b}). Subfigure \textbf{a}: temperature color code (dark red $\rightarrow$ red $\rightarrow$ yellow); Subfigure \textbf{b}: stable/unstbale branch color code (stable: red; unstable: blue).}
\label{fig6}
\end{figure*}
With a wide range of vorticity magnitudes, a three-dimensional surface can be created giving any output parameter as a function of the two input parameters, vorticity $(\omega)$ and ambient compressive strain rate $(S^*)$. Figure~\ref{fig6} shows two representations of these surfaces for maximum temperature and scalar dissipation rate by parabolic fitting of the three vorticity cases; $\omega = 0, 1, \sqrt{1.5}$. If viewed in the $T_{max}$-$S^*$ plane, Fig.~\ref{fig6}\textbf{a} appears as the black curves in Fig.~\ref{fig2}\textbf{a} and Fig.~\ref{fig6}\textbf{b} appears as Fig.~\ref{fig4}. These surface plots are given to illustrate the three-dimensional character of the rotational flamelet solutions. 

Convergence of the numerical scheme stipulates a less than one percent change in dependent variables between successive Newton iterations, except for minor species mass fractions which show greater volatility. Due to convergence difficulty especially near the extinction limit, the data in Fig.~\ref{fig2}-\ref{fig4} and Fig.~\ref{fig6}-\ref{fig8} has been post-processed to achieve smooth transitions between the stable and unstable branches. The associated error with these curve fits is less than $0.5 \%$.

\subsection{Unequal Transverse Strain Rates ($S_1 \neq S_2$) and Non-constant Lewis Number \label{subsec:ResultsSec2}} \addvspace{10pt}
The broader theory outlined in this manuscript is not limited to the assumption of unitary Lewis number or even constant Lewis number. The Fickian diffusion formulation assumed prior, can be replaced by the mixture-averaged or multicomponent diffusion formulation, as was done by Hellwig et al. \cite{Hellwig2024}. These are more accurate diffusion models in which the diffusivities for different species are unique and diffusion velocity depends on the gradients of multiple species. Both diffusion formulations are built on standard kinetic theory although simplifying ad hoc assumptions are made in the mixture-averaged form.
\begin{figure}[h!]
\centering
\includegraphics[width=250pt]{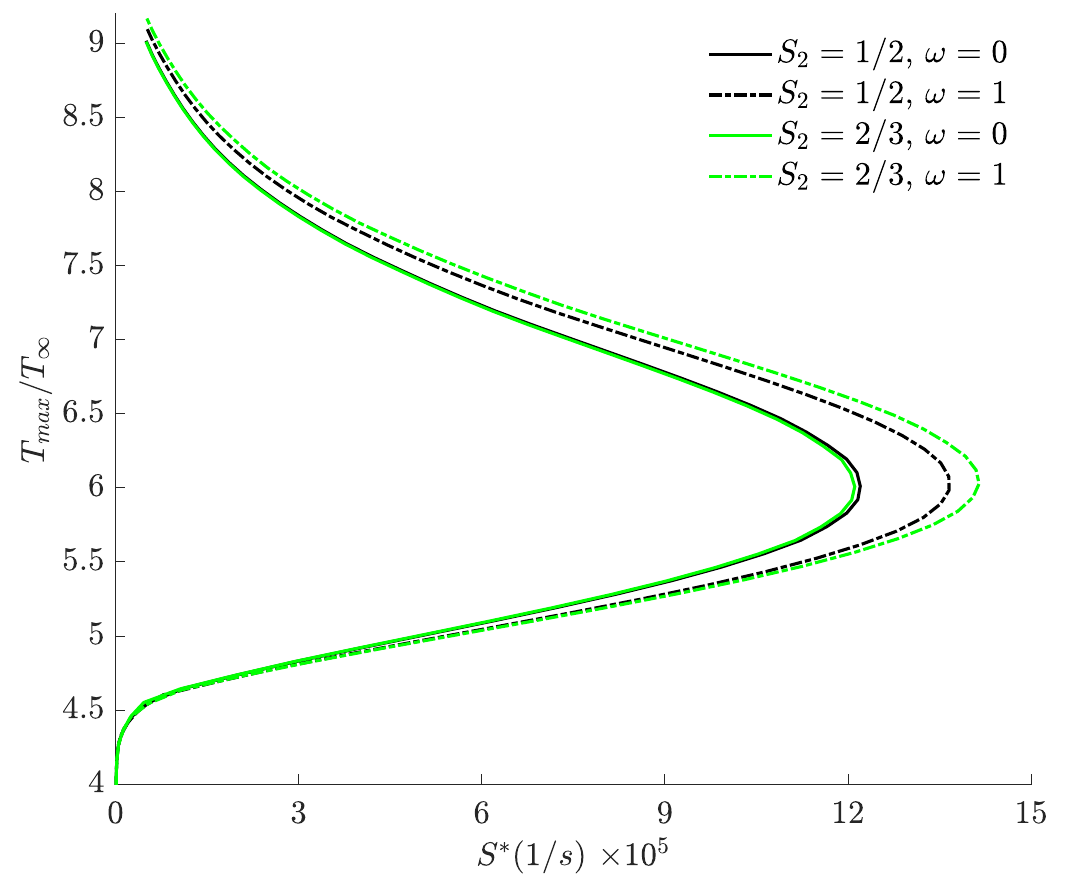}
\caption{\footnotesize S-curves using multicomponent diffusion with transverse ambient strain rate variation.}
\label{fig7}
\end{figure}

\begin{table*}[ht] \footnotesize
\centering
\caption{Output variables at constant transverse strain rates, $S_{1}^*$}
\begin{tabular}{ccccccc}
\hline
\text{$\omega$} & \text{$S_1, S_2$} & \text{$S^*$ $(1/s)$} & \text{$S_1 S^*$ $(1/s)$} & \text{$T_{max}$ $(K)$} & \text{$\int \dot{\omega}_{H_2O} d\eta$ $(kg/m^2s)$} & \text{$\chi_{max}$ $(1/s)$}  \\ \hline
0 & 1/2, 1/2 & 660,000   & 330,000 & 2010.5 &  6.4088 & 332,350 \\
0 & 1/3, 2/3 & 1,000,000 & 330,000 & 1817.1 &  6.8295 & 483,980 \\ 
1 & 1/2, 1/2 & 800,000   & 400,000 & 1996.0 &  6.5094 & 344,740 \\ 
1 & 1/3, 2/3 & 1,200,000 & 400,000 & 1813.2 &  6.8434 & 482,590 \\ \hline
\end{tabular}
\label{table1}
\end{table*}

Using a multicomponent or mixture-averaged diffusion formulation allows preferential and differential diffusion which breaks two commonly used practices in many flamelet theories: first, scalar dissipation rate is no longer clearly defined since the mass diffusivities of each species are different; secondly, counter-gradient diffusion becomes a possibility, especially if high molecular weight fuels are considered. In this case, the mixture fraction will not vary monotonically between the oxidizer and fuel inflows which prevents the traditional method of solving the problem in mixture fraction space. 
\begin{figure}[h!]
\centering
\includegraphics[width=250pt]{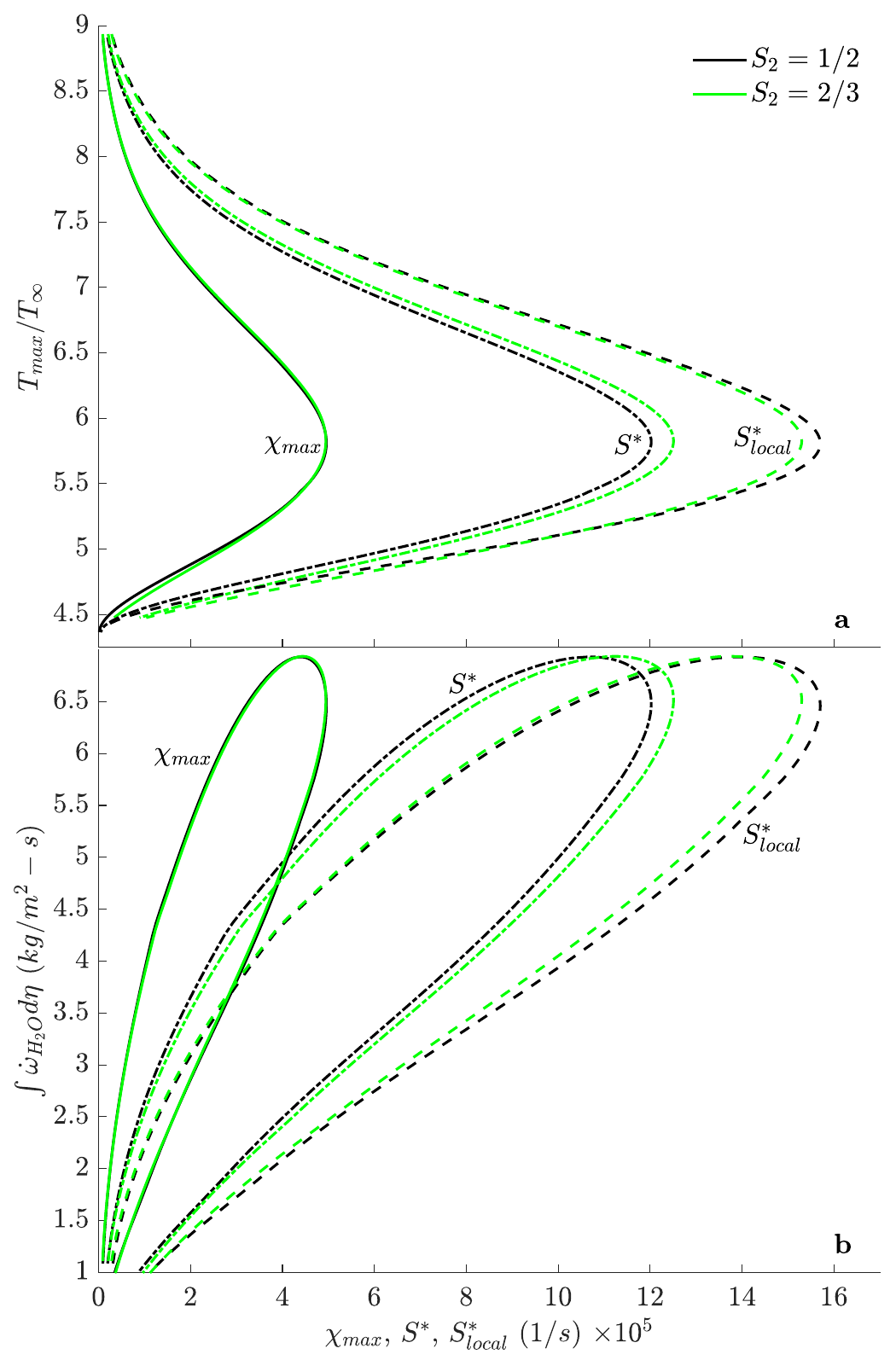}
\caption{\footnotesize Extinction curves versus: maximum scalar dissipation rate, $\chi_{max}$; ambient strain rate, $S^*$; maximum local strain rate, $S^{*}_{local}$. Vorticity is constant, $\omega = 1$, for all curves. $S_1 = 1/2,1/3$, $S_2 = 1/2,2/3$.}
\label{fig8}
\end{figure}

The data for Fig.~\ref{fig7} is taken from \cite{Hellwig2024} and shows S-curves similar to Fig.~\ref{fig2}\textbf{a} for $S_1 = S_2 = 1/2$ and $S_1 = 1/3$, $S_2 = 2/3$ using multicomponent diffusion. Of course, SDR does not appear on the abscissa since it is now undefined. Figure~\ref{fig8} shows a repeat of certain data in Fig.~\ref{fig2} ($Le = 1$); however, vorticity is fixed at $\omega = 1$, while $S_1$ and $S_2$ vary. Here again, $S_1 = S_2 = 1/2$ and $S_1 = 1/3$, $S_2 = 2/3$. These figures illustrate two points. First, the qualitative impact of vorticity on extinction limits is common among constant and variable Lewis number formulations. Secondly, the three-dimensional character of the inflows (magnitudes of $S_1$ and $S_2$) alters the extinction limits. With vorticity, a higher transverse strain rate aligned with the vorticity vector, $S_2$, yields a more stable flame for a given ambient compressive strain rate, $S^*$. This occurs because vorticity favors expulsion of low-density fluid outward in the direction unimpacted by centrifugal force, i.e., the direction with which $S_2$ and $\omega$ are aligned.

The original flamelet formulation, using SDR as formulated by Peters \cite{Peters1984} (see p. 181-184 of \cite{Peters2000}), applies a constant transverse strain rate in the $x$ direction. This strain rate corresponds to our $S_1$ only at $x = +\infty$ since our transverse strain rates vary with their respective axes and with $y$. Nevertheless, if two cases are compared in which the transverse strain rate in the $x$ directions are equal at $\infty$, that is $S_1S^*$ is the same value for two cases, we find peak temperatures differ by $\sim$10\%, burning rates differ by $\sim$5-6\%, and maximum SDR differs by $\sim$40-45\% as shown in Table \ref{table1}. Thus, it is clear that the assumptions of constant strain rates and potential flow omit certain physical behaviors that are possible to capture and should be captured.

\section{Conclusion\label{sec:unnum}} \addvspace{10pt}
The analysis is built on two hypotheses. First, the flamelet field within the counterflow is determined by the ambient scalar and velocity fields of the two incoming flows. This hypothesis is supported by the analytical results which show that the scalar properties, normal strain rates, and vorticity are influential parameters. To be precise, enstrophy ($\omega^{*2}$) is the parameter that appears in the analysis. Second, we believe these ambient properties are better suited for scaling relations with a resolved flow in a turbulent combustion analysis; however, any associated improvement to RANS or LES remains to be demonstrated.

Depending on these inflow conditions, the analysis shows particular three-dimensional spatial variations through the counterflow for the velocity and one-dimensional variation for the scalar properties: e.g., temperature, mass fractions, and SDR. Both a stable branch and an unstable branch are described with different property variations. Thus, two different spatial profiles and values of maximum temperature and integrated burning rate can appear for the same set of ambient conditions. However, nearly the same maximum scalar dissipation rate and the same maximum local strain rate or integrated burning rate can appear on both branches. The maximum temperature is a more distinctive single parameter in distinguishing between the two branches.

For a given set of input parameters, the introduction of vorticity and varying transverse strain rates significantly changes the velocity profiles and scalar gradients within the domain. Furthermore, the ambient strain rate at which the flame extinguishes may be increased as much as $26\%$ by vorticity. 

Existing flamelet theories avoid using vorticity and ambient inflow normal strain rates as critical input parameters \cite{Peters2000,PitschPeters1998,Mueller2020,Pitsch_Ihme}. Instead, together with pressure, local conditions at the flame are used for input to the analysis. Specifically, mixture fraction and SDR are used to identify all flame characteristics.

The identification of maximum SDR (usually assumed as a Gaussian variation) is claimed to allow identification of the important flame characteristics including production rate and maximum temperature in many existing flamelet models. While results show that the impacts of applied strain rates and vorticity are somewhat captured by maximum SDR, they also show that SDR and local strain rate depend strongly on vorticity and ambient inflow strain rate, including how strain rate distributes in a three-dimensional counterflow structure. Thus, SDR cannot be used to distinguish between all possible flamelet solutions of varying global inputs ($S^*$ and $\omega$) as multiple solutions appear the same under the guise of SDR. In addition, there is little literature that can guide the scaling of scalar gradients or SDR from the resolved scale for RANS or LES to the sub-grid scale of the flamelet. SDR (as commonly constructed) is essentially impossible to scale because it asymptotes to zero outside the flamelet. The scaling of vorticity or ambient (inflow) strain rate in turbulent flows is better understood and even is discussed in textbooks \cite{Hinze_1959,Pope_2000,Zhu_Antonia_1996,Johnson2024}. Finally, the use of SDR in flow media with preferential and differential diffusion is highly artificial and controversial. It is clear that a method relying on the scaling of inflow conditions for vorticity and normal strain rate is sound in its analytical foundations and in its ability to apply with models using more detailed transport; however, any realized improvements consequential of the former point remain to be seen.

Ultimately, the impacts of vorticity, transverse strain rates, and the authors decision to parameterize the flamelet solutions via scaled velocity derivatives, will be realized when the flamelet model is used in RANS and LES simulations and compared to CFD using other flamelet models and experiments. As of the time this paper was written, implementation of the new flamelet model in CFD simulations is an ongoing area of research. We look forward to publishing the results of this work so the community may have quantified metrics as to the differences in our model.

\newpage
\section*{Acknowledgements}

The research was supported by the Air Force Office of Scientific Research through Grant FA9550-22-1-0191 with Dr. Justin Koo as program manager and by the Office of Naval Research through Grant N00014-21-1-2467 with Dr. Steven Martens as program manager.

\newpage
\bibliographystyle{elsarticle-num-PROCI-jabbrv}
\bibliography{CNF_ZChi}

\end{document}